\documentclass[%
 reprint,
 amsmath,amssymb,
 aps,
 prl,
 nofootinbib
]{revtex4-2}

\usepackage{graphicx}
\usepackage{dcolumn}
\usepackage{bm}
\usepackage{physics}
\usepackage{enumerate} 
\usepackage[hidelinks]{hyperref} 
\usepackage{tikz}
\usepackage{xcolor}
\renewcommand{\div}{\nabla \cdot }
\renewcommand{\grad}{\nabla}

\begin{document}
\preprint{APS/123-QED}
\title{Spontaneous Ratchet Currents and Transition Dynamics in Active Wetting}

\author{Noah Grodzinski$^{1}$}\thanks{Corresponding author: njbg2@cam.ac.uk (he/him)}
\author{Robert L. Jack$^{1,2}$}
\author{Michael E. Cates$^{1}$}

\affiliation{$^1$Department of Applied Mathematics and Theoretical Physics, University of Cambridge, Wilberforce Road, Cambridge, United Kingdom}
\affiliation{$^2$ Yusuf Hamied Department of Chemistry, University of Cambridge, Lensfield Road, Cambridge, United Kingdom}
\date{\today}
\begin{abstract}

Self-propelled particles accumulate on repulsive barriers in so-called active wetting, whose relationship with equilibrium wetting remains unclear. Using an exact (noiseless) hydrodynamic framework for an active lattice gas, we show, using a slit geometry with periodic boundary conditions, that active matter exhibits both fully-wet and partially-wet states, with a critical wetting transition between them. Furthermore, we demonstrate the existence of a spontaneous-symmetry-breaking ratchet current in the partially-wet state, leading to departure of the bulk densities from their binodal values and the emergence of a novel dynamical pathway for the full-to-partial wetting transition. We elucidate this modified dynamical pathway using a minimal model. The results, while establishing a direct connection between active and equilibrium wetting, also identify the nonequilibrium consequences of activity.

\end{abstract}

\maketitle
\textit{Introduction--}
Active matter includes a broad class of self-propelled particle systems, locally consuming energy to produce directed motion \cite{Marchetti2013, Ramaswamy2010}. This leads to a range of intrinsically nonequilibrium behaviours, including motility-induced phase separation (MIPS) into liquid and vapour~\cite{Cates2015, Cates2025}. Examples of active systems range from flocking birds \cite{Vicsek1995, Toner1998} to swimming bacteria \cite{Tailleur2008, Elgeti2015}. Much of the collective behaviour in such systems depends weakly on the microscopic details and interactions \cite{Cates2013, Ramaswamy2010}. Therefore, particle-level \cite{Fily2012, Redner2013, Digregorio2018} and field-theoretic \cite{Wittkowski2014, Tjhung2018, Cates2025} models of active matter are natural settings for studying emergent collective behaviour.

Boundaries play a significant role in active systems. They can induce accumulation, drive steady-state currents, and even destroy bulk structure \cite{BenDor2022, Galajda2007, Reichhardt2017, Bechinger2016, Elgeti2015, Stenhammar2016, Granek2024, Mangeat2024}. {One prominent boundary-driven effect is the accumulation/adsorption of active particles on purely repulsive barriers \cite{Das2018, Elgeti2013, Sepulveda2017, Sepulveda2018, Caprini2024, Wysocki2020}, arising from the persistent self-propulsion of particles into the barrier. This process has become known as active wetting \cite{Thiele2026}, and is implicated in a variety of biological processes~\cite{Douezan2011, Joanny2013, Lauga2006, Chai2024, Harshey2003, PerezGonzalez2018, Brugues2014}.}

{In equilibrium systems, adsorption (a microscopically thin film of fluid near a surface) is carefully distinguished from genuine wetting transitions \cite{Binder1988, DeGennes1985, Sullivan1986, Dietrich1988, Schick1990, Bonn2009}. In particular, equilibrium critical wetting is a continuous surface phase transition on varying a surface attraction parameter at two-phase coexistence, at which the adsorbed film thickness diverges. A complete wetting transition involves the divergence of the wetting film, on approaching bulk coexistence from the wet side of the wetting transition. Simulations have suggested that in systems undergoing MIPS, accumulation of active particles \cite{Sepulveda2018, Sepulveda2017, Das2018, Elgeti2013} can develop into macroscopic wetting \cite{Turci2024, Zhao2026}, with transitions analogous to equilibrium complete wetting \cite{Neta2021} and critical wetting \cite{Turci2021, Das2020, PerezBastias2025, Das2025} (see also \cite{Grodzinski2026a}).} However, a unified theory of active and passive wetting has remained elusive, due to the confounding role of steady-state currents and 
active interfacial phenomena (which lack a single effective surface tension) \cite{Zhao2026, Bialke2015, Turci2024, Adkins2022, Granek2024, Zakine2020}.

In this Letter, we study active wetting by analysing a lattice model of self-propelled particles whose large-scale (hydrodynamic) behaviour can be derived exactly~\cite{KourbaneHoussene2018, Mason2023a, Erignoux2024}. The hydrodynamic limit suppresses noise, allowing study of active wetting in a context free of fluctuation-driven effects (so that critical behaviour is of mean-field type). We investigate wetting in a setup analogous to the ``slit geometry'' of classical wetting.

Using this hydrodynamic framework, we uncover two key results. First, active accumulation on barriers leads robustly to fully-wet {(or, completely-wet)} and partially-wet steady-states in the thermodynamic (many particle, large-scale) limit. While this has been observed previously in simulations \cite{Turci2024, Zhao2026, Neta2021, Turci2021, Das2020, PerezBastias2025, Das2025}, an exact description of these wetting states has not been available. {We also show that the transition between these wetting states shares phenomenology with equilibrium critical wetting (i.e. a continuous wetting transition on tuning the wall attraction at coexistence)}. This aspect of our work is discussed fully in the companion paper \cite{Grodzinski2026a}. Here we focus on our second key result, which is the demonstration of a ratchet effect in active wetting, arising due to spontaneously broken symmetry, as opposed to any asymmetry in the underlying geometry. We show that this drives an intrinsically nonequilibrium transition pathway between the full- and partial-wetting states.

Our results establish active wetting as a genuine surface phase transition, with an exact hydrodynamic description that directly parallels equilibrium {critical} wetting despite the absence of attractive inter-particle or particle-wall interactions. This generalises the analogy between bulk MIPS and equilibrium phase separation~\cite{Cates2015} to include surface phenomena, confirming that classical concepts of liquid-gas physics near surfaces remain relevant far from equilibrium. Simultaneously, the active nature of the particles manifests in \emph{qualitatively} new physics: in the slit geometry, spontaneous symmetry breaking produces circulating ratchet currents, modifying both the steady-states and the dynamical transition pathways. We expect such modified behaviour to be generic to systems that undergo MIPS.

\textit{Model--}
We study the active lattice gas introduced in~\cite{Mason2023a}. Particles move on an $N \times N$ square lattice, with at most one particle per site, in an external potential $V$; the lattice occupies an $L\times L$ domain. The hydrodynamic limit is $N\to\infty$ at fixed $L$, and fixed mean occupancy $\bar{\rho}.$ Each particle $i$ has a position $\vb{x}_i$ and an orientation vector $\vb{e}(\theta_i) = (\cos(\theta_i), \sin(\theta_i))$. The angle $\theta_i$ undergoes Brownian motion with diffusivity $D_O$, while the position $\vb{x}_i$ is governed by diffusion ($D_E$), self-propulsion (in the direction $\vb{e}_i$ at velocity $v_0$), and the external potential $V$. These processes have non-trivial dependence on $N$ such that both diffusion and activity remain relevant in the hydrodynamic limit \cite{KourbaneHoussene2018, Mason2023a}. The full microscopic dynamics, alongside a discussion of this hydrodynamic limit, are given in \cite{Grodzinski2026a}.  The system is described in the hydrodynamic limit by a probability density $f(\vb{x}, \theta, t)$ for particles with spatial position $\vb{x}$ and orientation $\theta$ at time $t$.
The local density (mean occupancy) and polarisation are defined as $\rho(\vb{x}, t) = \int_0^{2\pi} f(\vb{x}, \theta, t) \dd{\theta}\in [0, 1]$ and  $\vb{p}(\vb{x}, t) = \int_0^{2\pi} f(\vb{x}, \theta, t) \ {\vb{e}}(\theta) \dd{\theta}$ respectively.
The deterministic evolution of $f$ obeys:
\begin{equation}\label{eq:ABP_evolution}
  \pdv{f}{t} = - \div \vb{J}_\text{diff} - \text{Pe} \ \div \vb{J}_\text{activity} + \partial_\theta^2 f,
\end{equation}
where
\begin{align}
  &\vb{J}_\text{diff} = -d_s(\rho) \grad f - f \mathcal{D}(\rho) \grad \rho - f(1 - \rho) \grad V, \label{eq:J_diff}\\ 
  &\vb{J}_\text{activity} = (\mathcal{D}(\rho) - 1) f \vb{p} + \vb{e}(\theta) d_s(\rho) f. \label{eq:J_act}
\end{align}
The function $d_s(\rho)$ is accurately approximated by a polynomial (see \cite{Mason2023b, Mason2023a, Grodzinski2026a}); we define $ \mathcal{D}(\rho) = [1 - d_s(\rho)]/\rho$ for ease of writing.
The first two terms in $\vb{J}_\text{diff}$ were derived in \cite{Mason2023a}, while the final term (involving $V$) is derived in \cite{Grodzinski2026a}. While $D_E, D_O, v_0$ may be chosen arbitrarily, an equation of form (\ref{eq:ABP_evolution}) can always be obtained by rescaling $t \rightarrow D_O t$, $x \rightarrow x / \sqrt{D_E/D_O}$, such that the unit of time is the reorientation time $\tau \equiv D_O^{-1}$, and the length unit is the distance travelled by a particle due to diffusion in that time, $l_D \equiv \sqrt{D_E/D_O}$. The Péclet number, describing the strength of self-propulsion \cite{Mason2023a}, is $\text{Pe} \equiv \frac{v_0}{\sqrt{D_ED_O}} = v_0 \tau / l_D$.

\begin{figure*}[t!]
    \centering
    \includegraphics[width=\linewidth]{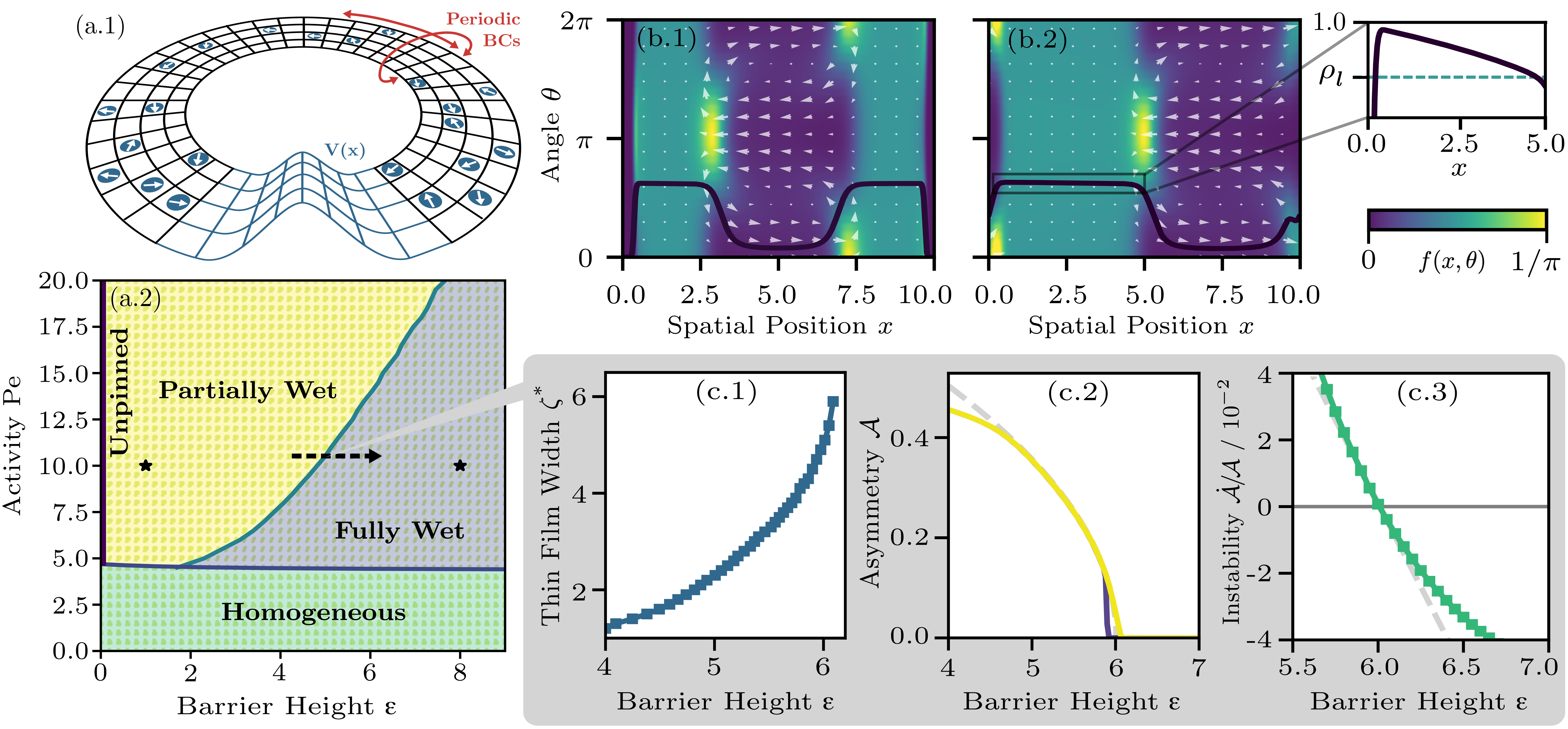}
    \caption{(a.1) Illustration of the system geometry. (a.2) Surface phase diagram; individual points are labelled according to their extrapolated steady-state. (b.1-2) Example full- and partial-wetting states; orientation-dependent density $f(x, \theta, t)$ is shown in colour; arrows show current (arbitrary scale); integrated spatial density $\rho(x, t)$ is shown with overlaid line. (c) Critical wetting phenomena. (c.1) Divergence of the thin film width. (c.2) Pitchfork bifurcation in the asymmetry, scanning $\mathcal A$ as a function of $\epsilon$ up (orange) and down (blue). (c.3) Linear stability of a fully-wet initial condition with weak noise; measured at early times, when asymmetry growth is approximately linear. See \cite{Grodzinski2026a} for further details of numerics. 
    }
    \label{fig:wetting_states}
\end{figure*}

Equilibrium wetting can be investigated in a \textit{slit geometry}, where a fixed mass of fluid at liquid-vapour coexistence is confined between planar walls \cite{Evans2019, VanLeeuwen1989, Sikkenk1987, Nijmeijer1990}. This allows investigation of wetting transitions without confounding effects such as curved interfaces, prewetting, or capillary condensation. 
To investigate active wetting, we choose an analogous geometry, inserting a planar barrier in a periodic geometry, at fixed total mass of active particles (as in~\cite{Turci2021, Das2020}). Since our model is deterministic and only one dimension does not have translational symmetry, densities depend only on one (periodic) spatial dimension $x$; our hydrodynamic results should apply in higher dimensions provided that planar MIPS interfaces are stable. As in the equilibrium analogue, fully-wet, partially-wet, and unpinned states correspond to macroscopic liquid layers on both sides, one side, and neither side of the barrier respectively (see \cite{Grodzinski2026a, Turci2021}). {Partially-wet states have only been observed in active systems on soft or penetrable walls \cite{Turci2021, Turci2024, Zhao2026, Das2020} --- these may model cell membranes or gel films separating fluid compartments. Therefore, to access the wetting transition,} and following \cite{Turci2021}, we study the case of a repulsive barrier with a smooth, finite potential $V(x) = \epsilon \left[ \cos\left( 2\pi x \right) + 1 \right]
H(1/2 - \abs{x})$ for Heaviside step $H(x)$. Our chosen geometry is illustrated in Figure \ref{fig:wetting_states}(a.1).

We vary two control parameters: the self-propulsion strength $\text{Pe}$, and barrier height $\epsilon$. For simplicity, the mean particle density is fixed at its critical value $\bar{\rho} = 0.659$ \cite{Mason2023a, Grodzinski2026a} throughout. The chosen system size $\ell_s = L/l_D$ is much larger than the typical interfacial width.

\textit{Wetting States--}
Figure \ref{fig:wetting_states}(a.2) shows the surface phase diagram, for a range of $\epsilon, \text{Pe}$. 
To obtain this, we initialise the system in a homogeneous state, with a small spatially uncorrelated perturbation. The system is then evolved until steady-state, numerically solving (\ref{eq:ABP_evolution}); this procedure is discussed further in \cite{Grodzinski2026a}. With a barrier present, above the critical activity, we observe fully-wet and partially-wet states; see Figure \ref{fig:wetting_states}(b.1-2). Unpinned (dry) states are not observed for $\epsilon > 0$, as expected; any barrier breaks the translational symmetry, localising the liquid layer in our noiseless dynamics.  The homogeneous state, in which phase-separation is not observed, occurs for $\text{Pe}< \text{Pe}^* = 4.33$. 

Alongside demonstrating the existence of full- and partial-wetting states, our hydrodynamic framework allows an investigation of the transition between them. We find that the four main features of a critical wetting transition (see \cite{Grodzinski2026a, Turci2021, DeGennes1985}) are all displayed in the active system, see Figure \ref{fig:wetting_states}(c.1-3). The first is a continuous divergence of the thin film width $\zeta^*$ (on the side of the barrier without a macroscopic liquid layer in the partial-wetting scenario). The second is a pitchfork bifurcation in the spatial asymmetry, 
$
    \mathcal{A} = \frac{1}{\ell_s(\bar{\rho} - \rho_v)}\int_0^{\ell_s/2} \rho(x) - \rho(\ell_s-x) \ \dd{x},
$
(whose random sign results from spontaneous symmetry breaking).
The third is a vanishing hysteresis loop; transitions from fully-wet to partially-wet states do not display metastability. Finally, the onset of linear instability of the fully-wet state occurs exactly at the transition, confirming absence of metastability. 

In the Cahn theory of equilibrium wetting \cite{Cahn1977}, critical wetting occurs when the wall significantly reduces the attractions between particles in its neighbourhood (see \cite{Cahn1977, DeGennes1985, Grodzinski2026a}). In the active wetting scenario, the large polarisation $\vb{p}$ means that active particles near the barrier are locally aligned. Since effective attractions in MIPS stem from collisions between oppositely aligned particles~\cite{Cates2015}, the barrier inhibits the attractive inter-particle interactions dramatically (absolutely so in the large-Pe limit). This explains physically our observation of a critical active wetting transition.

\textit{Ratchet Currents--}
The partially-wet state displays a steady-state ratchet current normal to the barrier, which flows from the ``dry'' side to the ``wet'' side, circulating around the periodic domain. Although the ratchet principle predicts such currents for any system of (non-momentum-conserving) active particles in an {\em asymmetric} potential \cite{Reichhardt2017, Metzger2025}, the current here is a result of spontaneously broken mirror symmetry about a \textit{symmetric} barrier. 
This is also distinct from \textit{flocking} behaviour \cite{Toner1998}, for which particles must have aligning interactions.

We define the \textit{density} current $J^{(\rho)}(x)$ such that $\partial_t \rho = - \partial_x J^{(\rho)}$.
Integrating (\ref{eq:ABP_evolution}) over orientations $\theta$ yields:
\begin{equation}\label{Eq:density_dynamics}
    J^{(\rho)}(x) = - [\partial_x \rho + \rho (1-\rho) \partial_x V ] + \text{Pe} (1-\rho) p_x
\end{equation}
This current must be a constant in steady-state; using the steady-state identity $\int_0^{\ell_s} \vb{p}(x) \ \dd x = \vb{0}$, we obtain (see Appendix):
\begin{equation}\label{eq:SS_current}
    J_{\mathrm{ss}}^{(\rho)} = -M[\rho] \int_0^{\ell_s} \rho(x) \pdv{V}{x} \ \dd x,
\end{equation}
where the effective mobility 
$
M[\rho] = \{\int_0^{\ell_s} [1-\rho(x)]^{-1} \dd x\}^{-1}
$
is strictly positive. From (\ref{eq:SS_current}), we see that current flows from the dry (low-density) side of the barrier to the wet (high-density) side, which is in the opposite direction to a diffusive flux through the barrier.

For large systems $\ell_s \to \infty$, the mobility decays as $M \sim \ell_s^{-1}$; the integrand in \eqref{eq:SS_current} is non-zero only for $|x|< 1/2$, so the integral remains of order unity.  Hence the current is suppressed in large systems as 
$
J_{\mathrm{ss}}^{(\rho)} \sim \ell_s^{-1} \; .
$
Physically, this scaling arises because the ratchet acts as a localized driving force, while the friction is distributed across all particles.

\begin{figure}[h!]
    \centering
    \includegraphics[width=1.0\linewidth]{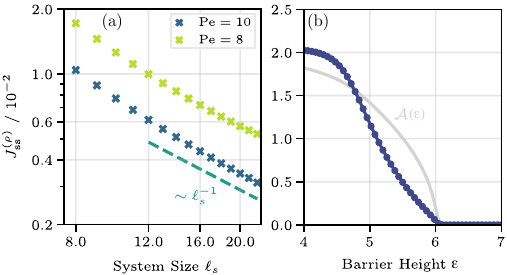}
    \caption{Illustration of steady-state current in the partially-wet state. (a) Steady-state current as a function of system size, tending to $\sim1/\ell_s$. (b) Steady-state current vanishes continuously as $\epsilon \to \epsilon^*$ (same parameters as Figure \ref{fig:wetting_states}(c.2)).}
    \label{fig:current}
\end{figure}

The system-size dependence of $J^{(\rho)}_\text{ss}$ is illustrated in Figure \ref{fig:current}(a). Alongside the expected asymptotic scaling, current flows in the predicted direction and is uniform to numerical accuracy. The steady-state current is also observed (see Figure \ref{fig:current}(b)) to vanish continuously as $\epsilon \to \epsilon^*(\rm Pe)$, as expected for a critical transition. The numerical solutions reveal moreover that the steady-state density profile in the partially-wet state is modified, with the liquid layer developing a linear density gradient across its bulk, see Figure \ref{fig:wetting_states}(b.2, magnified panel). A similar gradient develops in the vapour phase, but is less obvious as the tail of the liquid-vapour interface into the vapour bulk is more pronounced (as $d_s(\rho_v) > d_s(\rho_l)$). 
As the gradients must scale inversely with system size, and the bulk layers scale directly with  system size, we expect an asymptotically constant density drop across the bulk layers. {Hence, the ratchet current produces an $\mathcal{O}(1)$ departure of the bulk densities from their binodal values that persists as $\ell_s \to \infty$, as a non-local consequence of the wet barrier.  There is no such bulk effect in equilibrium wetting.}

The ratchet current identified in our framework is qualitatively different to that discussed in \cite{Zhao2026}, which showed how a \textit{local} circulating current around the contact line modifies the contact angle in active wetting droplets. In our case, activity induces a spontaneous \textit{global} current through a permeable barrier, with the consequences below. 

\begin{figure}[h]
    \centering
    \includegraphics[width=1.0\linewidth]{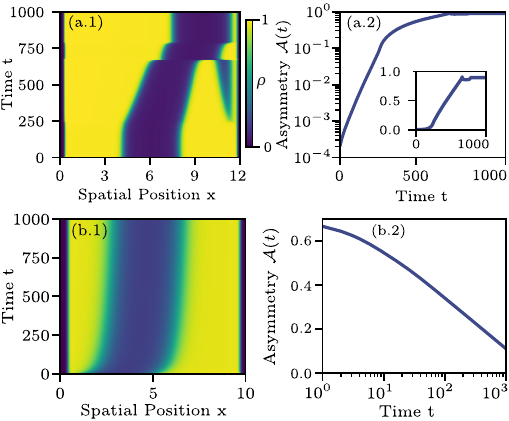}
    \caption{(a.1) Full-to-partial transition pathway at $\text{Pe}=15$. (a.2) $\mathcal{A}(t)$ for this process. (b.1, b.2) Pathway and asymmetry for partial-to-full transition at $\text{Pe}=8$.}
    \label{fig:dynamical_pathways}
\end{figure}

\textit{Dynamical Transitions--} The ratchet current has important dynamical effects, despite being suppressed in large systems. As shown below, it modifies the \textit{dynamical transition pathway} from fully-wet to partially-wet states, thereby qualitatively changing the wetting behaviour of active systems compared to their passive counterparts. 

To dynamically observe a full-to-partial wetting transition, we initialise the system in a fully-wet state with $\epsilon>\epsilon^*$, then lower $\epsilon$ to below the critical barrier height such that a partially-wet state becomes the (only) stable stationary solution to (\ref{eq:ABP_evolution}). We observe that the transition pathway then proceeds in three distinct stages (see Figure \ref{fig:dynamical_pathways}(a)):

\textit{(i) Linear instability about the barrier.} A small initial asymmetry about the barrier is amplified by the anti-diffusive ratchet current, leading to an exponentially growing asymmetry,
$\mathcal{A}(t) = \mathcal{A}_1 e^{\lambda (\epsilon^* - \epsilon) t} $  (up to time $t_1 \simeq 250$, independent of liquid-layer width).

\textit{(ii) Evaporation of one liquid layer.} The ratchet-driven instability induces a depression in density on one side of the barrier, causing its liquid layer to ``pinch off'' and form a droplet. This droplet then evaporates at approximately constant rate (see below), leading to a linear increase in the asymmetry, 
$
    \mathcal{A}(t) \simeq \mathcal{A}_2 t
$ (up to time $t_2 \simeq 650$).

\textit{(iii) Bursting and saturation.} The liquid droplet then ``bursts", redistributing its mass onto the other liquid layer and the boundary. Soon afterward (at time $t_3 \simeq 800$), a saturated partially-wet state is reached with constant asymmetry, $\mathcal{A}(t) = \mathcal{A}_3$.

After stage (i) described above, the asymmetric density profile on the barrier is very weakly dependent on time, leading to an approximately constant current through the barrier. This ratchet current redistributes mass from the detached liquid layer to the other (wet) side, meaning that layer evaporates in a time approximately linear in its width. 

In contrast, the partial-to-full wetting transition (like the passive counterpart) proceeds via interaction between the exponential tails of the liquid-vapour interface and the barrier. This process becomes extremely slow in the late, distant-interface, stage of the transition, see Figure \ref{fig:dynamical_pathways}(b.2). The ratchet-induced dynamical pathway therefore provides a \textit{much faster} mechanism for full-to-partial wetting transitions than any passive counterpart, especially in large systems. Indeed, a smaller system size and a smaller value of Pe are required in Figure \ref{fig:dynamical_pathways}(b) for that transition to occur in a reasonable time.

\textit{Minimal model--} To further understand this ratchet-driven dynamical pathway, we seek a minimal model which captures the ratchet dynamics while excluding other active effects. A derivation of this model as an approximation of the full dynamics is contained in the Appendix. We consider a scalar field $\phi: [-\ell_s/2, \ell_s/2) \rightarrow \mathbb{R}$, whose dynamics is given by:
\begin{equation}\label{eq:minimal_model_dynamics}
    \pdv{\phi}{t} = - \partial_x \left(  \partial_x  \qty(\frac{\delta F^\phi_\text{PS}}{\delta \phi}) + \eta \, \delta(x) \, \partial_x \phi  \right)\,.
\end{equation}
Here the first term represents passive Cahn-Hilliard phase separation, and the second an active pump at the origin with strength $\eta$, which induces a current with the same system-size scaling as (\ref{eq:SS_current}). The free energy is of Cahn-Hilliard type with minima at $\phi_{v,l} = 0, 1$:
\begin{equation}\label{eq:minimal_F_PS}
    F^{\phi}_{\mathrm{PS}} = 
\int_{-\ell_s/2}^{\ell_s/2} \! \dd{x} \,
\left\{
\frac{\kappa}{2} \left( \partial_x \phi \right)^2 
+ \frac{\alpha}{4} \phi^2 (\phi - 1)^2
\right\}.
\end{equation}

In this model, we can solve for the linear instability of a bulk liquid domain of length $\ell_L$ which contains at its centre a pump-- modelling the full-wetting state. This instability, which always leads to a partial-wetting state (with the liquid bulk on only one side of the origin), occurs for $\eta > \eta^*$ (see Appendix), where $\eta^* = \sqrt{2 \alpha \kappa} + \mathcal{O}(\ell_L^{-1})$. 
\begin{figure}[t]
    \includegraphics[width=\linewidth]{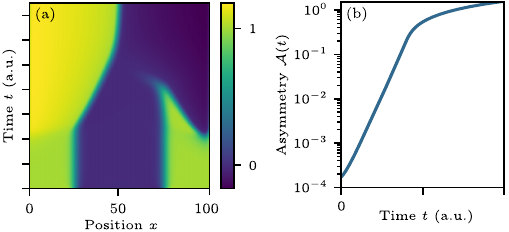}
    \caption{(a) Density profile over time for a full-to-partial transition in the minimal model ($\alpha = \kappa = 1, \eta = 2$). (b) Asymmetry over time. Comparing with Figure \ref{fig:dynamical_pathways}(a), one sees that the minimal model captures the three-stage transition found in the full dynamics.}\label{fig:minimal_model}
  \end{figure}
Since $\eta^*$ does not diverge as $\ell_L \to \infty$,  a finite localised pump can trigger a wetting transition even in the large-system limit. 
Additionally, the fact that $\eta^*$ is non-vanishing implies that passive physics is sufficient to prevent a transition to the partially-wet state for a weak pump. Overall, the full-to-partial dynamical pathway in the minimal model, as shown in Figure \ref{fig:minimal_model}, appears very similar to the full dynamics, even though the barrier and fluid are greatly simplified.

\textit{Conclusion--} We have shown that active wetting on repulsive barriers is a genuine surface phase transition, directly analogous to equilibrium wetting. Using an exact hydrodynamic framework, we obtain  steady states and demonstrate that fully- and partially-wet configurations, alongside a critical wetting transition, persist robustly in the thermodynamic limit. Our Letter therefore extends the analogy between MIPS and equilibrium phase separation to include surface phenomena.

At the same time, we uncover  intrinsically nonequilibrium behaviour. The partially-wet state exhibits spontaneous symmetry breaking, leading to a circulating ratchet current. (While in one spatial dimension this requires periodic boundaries, we expect a similar outcome, albeit with a more complicated current pattern, in higher dimensions under general distant boundary conditions.)  This current causes a departure of the bulk densities from their binodal values, and drives a qualitatively new transition pathway between wetting states which proceeds much faster than any passive analogue. We expect this behaviour to be generic to active wetting, and evident even in the large-system limit. These results reveal active wetting as a fertile setting in which equilibrium-like surface transitions and intrinsically nonequilibrium effects meet.

Recent advances in colloidal systems \cite{Palacci2014, Buttinoni2013, FinsCarreira2024} suggest several promising routes for experimental tests of these findings. For example, light-activated Janus particles can be confined within quasi-one-dimensional channels, in circular/toroidal geometries where periodicity is enforced by design. Inserting a permeable barrier \cite{Bechinger2016} in such a setup could experimentally test the existence of active wetting states, and the nonequilibrium effects predicted above.

\textit{Acknowledgments--} We thank Robert Evans, Francesco Turci, Maria Bruna, {Nigel Wilding, and Uwe Thiele} for insightful discussions and valuable feedback throughout this work. This work was supported by Engineering and
Physical Sciences Research Council under Grants No.\ EP/W524633/1 
(Project Reference 2927750) and No.\ EP/Z534766/1.

\textit{Data Availability--} The code and data that support our results are openly available at~\cite{Grodzinski2026b}.

\bibliography{apssamp}

\section{APPENDIX}
\subsection{Derivation of the Ratchet Current}\label{app:ratchet_current}
We derive \eqref{eq:SS_current} for the steady state density current, working from the dimensionless dynamics (\ref{eq:ABP_evolution}). We start from the expression for the spatial probability current in (\ref{eq:J_diff}, \ref{eq:J_act}), for which only the x-component is non-vanishing:
\begin{multline}
  J_x(x, \theta) = - \left[ d_s(\rho) \partial_x f + f \mathcal{D}(\rho) \partial_x \rho + (1-\rho) f \partial_x V \right] \\ + \text{Pe} \left[  (\mathcal{D}(\rho)-1) f p_x + \cos(\theta) f d_s(\rho) \right] \; 
\end{multline}
Integrating over $\theta$ and making use of the definition of $\mathcal{D}(\rho)$, we obtain the spatial density current 
\begin{equation}
    J^{(\rho)} =  -\left[ \partial_x \rho + (1-\rho)\rho \partial_x V \right] + \text{Pe} (1-\rho) p_x \label{eq:appendix_density_current}
\end{equation}

In steady state, the density current must be divergence-free, and so its one non-vanishing component must be a constant $J^{(\rho)}(x) = J^{(\rho)}_{\text{ss}}$. Dividing (\ref{eq:appendix_density_current}) through by $1-\rho$, integrating over space with periodic boundaries, and using the identity $\int_0^{\ell_s} \vb{p} \ \dd x = 0$ in steady-state\footnote{Orientational dynamics is independent of position, so the total polarisation must generically vanish. This can be proven easily from the dynamics.}, yields:
\begin{equation}
   J^{(\rho)}_{\text{ss}} = M[\rho] \int_0^{\ell_s} V(x) \, \partial_x \rho \dd{x} \; 
   \label{equ:J-MM-parts}
\end{equation}
where $M[\rho] = [ \int_0^{\ell_s} \frac{1}{(1-\rho)} \dd x]^{-1}$.  Integration by parts on \eqref{equ:J-MM-parts} yields \eqref{eq:SS_current}.

\subsection{Derivation of Minimal Model}\label{app:minimal_model}

We motivate the minimal model as an approximation to the dynamics of the active lattice gas.

Starting with the dynamics of the density field in the full model and integrating \eqref{eq:ABP_evolution} over the particles' orientations $\theta$ yields
\begin{equation}\label{eq:dynamics_mobility}
    \partial_t \rho({x}) = - \partial_x [{J}^{(\rho)}({x})] \equiv- \partial_x \qty[ \mathcal{M}(\rho({x})) \mathcal{F}({x}) ]
\end{equation}
where $\mathcal{M}(\rho) = 1-\rho \geq 0$ plays the role of a mobility. The generalised force $\mathcal{F}({x})$ can be split into two contributions:
\begin{equation}\label{eq:gen_forces}
    \mathcal{F} = \overbrace{\partial_x \ln(1 - \rho) + \text{Pe} \, p_x}^{\mathcal{F}_\text{PS}} - \overbrace{\rho \, \partial_x V}^{\mathcal{F}_\text{barr}}
\end{equation}
where the phase-separating force $\mathcal{F}_\text{PS}$ gives rise to MIPS, and $\mathcal{F}_\text{barr}$ arises from the barrier. To simplify this expression, we first seek a closed-form expression for the polarisation $p_x$.
We first use the definition of $\vb{p}$ to write   
\begin{multline}
    \partial_t p_i = \div \qty[ d_s(\rho) \grad p_i + \mathcal{D}(\rho) p_i \grad \rho + (1-\rho) p_i \grad V ] \\
    - \text{Pe}\ \div \qty[ (\mathcal{D}(\rho)-1) p_i \vb{p}] -\text{Pe}\ \sum_j \partial_j \qty[\frac{d_s(\rho)}{2} Q_{ij} ]\\
    -\text{Pe}\ \partial_i \qty[\frac{d_s(\rho)}{2} \rho] - p_i,
\end{multline}
where the nematic alignment tensor is $Q_{ij} (\vb x, t) = 2 \int f(\vb x, \theta, t) (\vb{e}_\theta)_i (\vb{e}_\theta)_j \dd \theta - \rho(\vb x,t) \delta_{ij}$.

Focusing on regions away from the barrier, we seek an effective theory for the density alone.  To this end, we make three approximations: (i) the polarisation $\vb{p}$ relaxes faster than the density $\rho$ allowing an adiabatic elimination; (ii) density gradients are not too large, which enables gradient expansion; (iii) particles' alignment is weak, which means $|\bm{p}|^2 \ll 1$ and $\underline{\underline{Q}}\approx 0$.  This last approximation is natural, given the previous two.

Next, considering regions away from the barrier we use weak alignment, and adiabatic elimination of the ``fast'' field $\vb{p}$, to obtain:
\begin{equation}\label{eq:p_implicit}
    p_i = \div \qty[ d_s(\rho) \grad p_i + \mathcal{D}(\rho)p_i \grad \rho] - \frac{\text{Pe}}{2} \partial_i [\rho \, d_s(\rho)] 
\end{equation}

Plugging (\ref{eq:p_implicit}) into itself recursively, and retaining only terms up to third-order in the gradient, yields
\begin{multline}\label{eq:p_explicit_app}
    p_i = - \frac{\text{Pe}}{2} \div \bigg[ d_s(\rho) \partial_i \grad (\rho \, d_s(\rho)) \\ + \mathcal{D}(\rho) \partial_i (\rho \, d_s(\rho))\grad \rho \bigg] - \frac{\text{Pe}}{2} \partial_i [\rho \, d_s(\rho)]
\end{multline}
Physically, the main effect of the third-order gradient terms on the dynamics is to prevent density gradients from growing too large.  The same effect can be accomplished in a simpler way by approximating the polarisation as
\begin{equation}
    p_x = -\frac{\text{Pe}}{2} \partial_x [\rho \, d_s(\rho)]  + \kappa_0 \, \partial_x^3 \rho
\end{equation}
with constant $\kappa_0>0$. Substituting this expression into the phase-separating force $\mathcal{F}_\text{PS}$ from \eqref{eq:gen_forces}, and defining $\kappa = \text{Pe} \, \kappa_0$ yields:
\begin{equation}
    \mathcal{F}_\text{PS} = \partial_x \qty[ \ln(1-\rho) - \frac{\text{Pe}^2}{2} \rho \, d_s(\rho)] + \kappa\,  \partial_x^3 \rho
\end{equation}
We recognise $\mathcal{F}_\text{PS}$ as the gradient of a free energy $F_\text{PS}$:
\begin{align}
    \mathcal{F}_\text{PS} &=- \partial_x \qty(\frac{\delta F_\text{PS}}{\delta \rho}) \\
    F_\text{PS} &= \int_0^{\ell_s} \dd{x} \qty{ \frac{\kappa}{2} (\partial_x \rho)^2 + f_\rho(\rho) }
\end{align}
where the bulk free energy density is
\begin{equation}
    f_\rho(\rho) = \int \frac{\text{Pe}^2}{2} \rho \, d_s(\rho) - \ln(1-\rho) \dd{\rho}
\end{equation}
This function becomes a double-welled potential for $\text{Pe} > \text{Pe}^*$, giving rise to MIPS. 

We now seek a scalar field in one dimension $\phi(x)$ whose dynamics minimally recreates the behaviour of $\rho$, with the same periodic boundary conditions. We simplify the full dynamics by replacing $f_\rho$ with a simple double-well potential $f_\phi = \frac{\alpha}{4} \phi^2 (1-\phi)^2$. The resulting free energy of the minimal model is then \eqref{eq:minimal_F_PS}.

To model the barrier within the minimal model, we only include its ratchet effect.  To this end,
we replace the potential $V$ with a simplified form $V(x) = \eta \, \delta(x)$ in $\mathcal{F}_\text{barr}$, yielding
\begin{align}
    \mathcal{F}_\text{barr}(\vb{x}) &= -\eta \, \rho(\vb{x}) \grad \delta(x) \\
    &=  -\eta \grad \qty[\rho(\vb{x}) \, \delta(x)] + \eta \, \delta(x) \grad \rho \,  . \label{eq:F_barr_two_terms}
\end{align}
The first term of \eqref{eq:F_barr_two_terms} models 
the effect of volume exclusion from the barrier. This does not contribute to the steady-state current overall as it takes the form of a total derivative, and so its integral over space vanishes. We therefore drop this term and retain only the second term of \eqref{eq:F_barr_two_terms} in the minimal model, identifying this as a pumping (ratchet) effect of the barrier. The resulting force in the minimal model is 
\begin{equation}
    \mathcal{F}^\phi_\text{pump}(x) = \eta \, (\partial_x \phi) \, \delta(x).
\end{equation}
To complete the minimal model, we replace the density-dependent mobility $\mathcal{M}(\rho)$ of \eqref{eq:dynamics_mobility} with a constant (taken without loss of generality to be $1$). 
We finally arrive at the minimal model \eqref{eq:minimal_model_dynamics}. This model replaces MIPS with Cahn-Hilliard phase-separation, while the effect of the barrier is reduced to a local pumping effect at the origin; volume exclusion and all other consequences of the barrier have been removed.

We also note that the momentum injection term in the minimal model results in a steady-state current with the same scaling as the full dynamics, in the limit of a large system. 

\subsection{Onset of Instability in the Minimal Model}
As noted above, the fully-wet state always exists as a steady-state solution of the minimal dynamics.  However, this state may not be linearly stable.   The critical wetting transition corresponds to the onset of linear instability of this state, just as in equilibrium \cite{Cahn1977, Grodzinski2026a}.  

We find this instability by considering a small perturbation of a homogeneous liquid state.   We substitute a periodic, mass-conserving perturbation
$
    \phi(x, t) = \phi_l + \sum_{n \geq 1} A_n \sin(q_n x) + B_n \cos(q_n x)
$
into the dynamics, where $q_n=2\pi n/L$; the even cosine modes play no role, and we obtain at leading order $  \dot{A}_n = \Gamma_{nm} A_m$, with 
\begin{align}
   \Gamma_{nm} = - \left( \frac{\alpha}{2} q_n^2 + \kappa q_n^4 \right) \delta_{nm} + \frac{2\eta}{L} q_n  q_m 
\end{align}

The (infinite) matrix $\Gamma$ has the form
$
    \Gamma = D + \beta \vb{q} \vb{q}^T
$, 
where $D = \text{diag}(\{ -\frac{\alpha}{2} q_n^2 - \kappa q_n^4 \})$ and $\beta = \frac{2\eta}{L}$. All the eigenvalues of $D$ are negative (as $\alpha, \kappa \geq 0$).
The Sherman-Morrison formula \cite{Sherman1950} states, for a rank-1 update of a diagonal matrix:
\begin{equation}
    \det(D + \beta \vb{q} \vb{q}^T - \lambda I) = \det(D- \lambda I)( 1 + \beta \vb{q}^T (D- \lambda I)^{-1} \vb{q} ).
\end{equation}
The eigenvalues of $\Gamma$ can therefore be obtained by finding the roots of its characteristic polynomial, $\det(\Gamma-\lambda I)$.  Using that $D$ is diagonal we obtain:
\begin{align}
    0 & = \det(D- \lambda I) \left( 1 + \beta \sum_{m} \frac{q_m q_m}{d_m - \lambda} \right) 
\nonumber\\
    & = \prod_n \left( d_n - \lambda \right) \,  \left( 1 - \beta \sum_{m} \frac{q_m^2}{\frac{\alpha}{2} q_m^2 + \kappa q_m^4 + \lambda} \right)
\end{align}
where $d_m = -\frac{\alpha}{2} q_m^2 - \kappa q_m^4$ are the eigenvalues of $D$. 

The matrix $\Gamma$ is real-symmetric so its eigenvalues are real.  Hence the boundary of linear stability occurs when this matrix has a single zero eigenvalue, with all others being negative.  Setting $\lambda=0$ and noting that $d_n<0$ for all $n$ we find
$
0 = 1- \beta \sum_{m} \frac{1}{ (\alpha/2) + \kappa q_m^2}
$. Using the identity $\sum_{n=1}^\infty \frac{1}{n^2 + a^2} = \frac{a \pi \coth(a \pi) - 1}{2a^2}$ (where $a = \frac{L}{2 \pi}\sqrt{\frac{\alpha}{2 \kappa}}$) and rearranging,  we obtain an explicit expression for the critical pumping strength $\eta^*$ at which the system becomes unstable:
\begin{equation}
    \eta^* = \frac{\sqrt{2 \alpha \kappa}}{\coth(L \sqrt{\frac{\alpha}{8 \kappa}}) - \frac{1}{L}\sqrt{\frac{8 \kappa}{\alpha}}}
\end{equation}
For large $L$, this tends to $\sqrt{2\alpha\kappa}$. Simulations confirm this instability boundary represents the onset of partial wetting, even when the ratchet is embedded in a liquid layer of finite width $L$ surrounded by vapour (instead of a periodic domain of the liquid density).

\end{document}